\documentclass[fleqn,10pt]{wlscirep}
\usepackage[utf8]{inputenc}
\usepackage[T1]{fontenc}
\title{Theory of optical tweezing of dielectric microspheres in chiral host media and its applications }

\usepackage{color}
\usepackage{amsmath}
\usepackage{ulem}
\usepackage{gensymb}
\usepackage{mwe}
\usepackage{bbding}
\usepackage{mathtools,amssymb,lipsum}

\usepackage{amsmath}

\newcommand{\be}{\begin{equation}}
\newcommand{\ee}{\end{equation}}
\newcommand{\beq}{\begin{eqnarray}}
\newcommand{\eeq}{\end{eqnarray}}
\newcommand{\bea}{\begin{array}}
\newcommand{\eea}{\end{array}}

\author[1,3,\Envelope]{Rfaqat Ali }
\author[2]{Rafael S. Dutra}
\author[1]{Felipe  A. Pinheiro}
\author[1]{Felipe S. S. Rosa}
\author[1]{Paulo A. Maia Neto}
\affil[1]{Instituto de F\'isica, Universidade Federal do Rio de Janeiro, Caixa Postal 68528, Rio de Janeiro, RJ, 21941-972, Brasil}
\affil[2]{LISComp-IFRJ, Instituto Federal de Educa\c{c}\~ao, Ci\^encia e Tecnologia, Rua Sebasti\~ao de Lacerda, Paracambi, RJ, 26600-000, Brasil}
\affil[3]{Photonics Research Center, Applied Physics Department, Gleb Wataghin Physics Institute, P.O. Box 6165,
University of Campinas - UNICAMP, 13083-970 Campinas, SP, Brazil}
\,
\affil[\Envelope]{Corresponding Author:  rali.physicist@gmail.com }

\begin{abstract}
We report for the first time the theory of optical tweezers of spherical dielectric particles embedded in a chiral medium.
We develop a partial-wave (Mie) expansion to calculate the optical force acting on a dielectric microsphere illuminated by a circularly-polarized, highly focused laser beam.  
When choosing a polarization with the same handedness of the medium, the axial trap stability is improved, thus allowing for tweezing of high-refractive-index particles.
When the particle is displaced off-axis by an external force, 
its equilibrium position is rotated around the optical axis by the mechanical effect of an optical torque. 
Both the optical torque and the angle of rotation are greatly enhanced in the presence of a chiral host medium when
considering radii a few times larger than the wavelength. In this range, the angle of rotation depends strongly on the microsphere radius and the chirality parameter of the host medium, opening the way for a quantitative characterization of both parameters.
Measurable angles are predicted even in the case of naturally occurring chiral solutes, allowing for a novel all-optical method to locally probe the chiral response at the nanoscale.
\end{abstract}
\begin{document}

\flushbottom
\maketitle
%
%
\thispagestyle{empty}

One of the landmarks in the field of optomechanics was the advent of optical tweezers \cite{ashkin1986},  that 
allows for manipulation of microsized dielectric spheres and nanostructures trapped by a 
 single tightly focused laser beam. Several applications in cell  \cite{ashkin1987a,Thalhammer2011,Araujo2012} and molecular biology \cite{Bustamante2011,fazal2011}, chemistry
 \cite{Yodh2001}, nanotechnology~\cite{Marago2013} and  physics \cite{Brut2012,Martinez2015,Arzola2017,Rosales-Cabara2020} have been developed 
  (for reviews see \cite{Grier2003,Ashkin2006,Polimeno2018,Gieseler2020}). 
 The physical phenomena behind the operation of optical  tweezers rely on momentum conservation as the incident trapping beam interacts with the microsphere.
 On one hand, within the geometrical optics approximation, the refracted light rays provide the key contribution to the optical force, which points towards the
 focal point, whereas reflected rays provide a detrimental 
 radiation pressure contribution \cite{ashkin1992}. On the other hand, the Mie-Debye theory  accounts for the exact wave-optical redistribution  \cite{Neto2000,Mazolli2003,Dutra2014} of linear momentum 
 engendered by Mie scattering at the trapped microsphere. 
 Playing with directional scattering in a metamaterial platform allows for trapping of high-index microspheres~\cite{rali2018}.

In addition to linear momentum, a circularly-polarised (CP) light beam also carries spin angular momentum (SAM) 
that can be transferred to trapped particles~\cite{Friese1998,Bishop2003,Ruffner2012,Yan2020}. 
The resulting optical torque leads to a rotation of the equilibrium position when a Stokes drag force is simultaneously applied~\cite{Diniz2019}. 
The optical force on chiral particles has been employed for chirality sorting and recognition in optical traps  \cite{Tkachenko2014,hayat2015,Schnoering2018,Shi2020}.
Enantioselective optical manipulation schemes have been proposed~\cite{Bradshaw2015,Chen2016,Zhao2016,Patti2019}, including the possibility of employing optical torques~\cite{Zhang2019,rali2020}.
The  transfer of  optical angular momentum dramatically changes when trapping particles  
in a chiral host medium~\cite{Zhang2018}.

In this paper, we develop the theory of optical tweezers in a chiral medium. Our approach is based on the full Mie scattering solution 
for spherical particles embedded in a chiral medium \cite{Hinders1992}, combined with a Debye-type non-paraxial vector model for the trapping beam \cite{RichardsWolf1959}. 
We consider a CP trapping beam and calculate the optical torque resulting from the transfer of SAM to the trapped particle. 
When a lateral Stokes drag force is simultaneously applied \cite{Diniz2019,rali2020},
the equilibrium position is displaced sideways and rotates around the optical axis as a consequence of the optical torque. 
The angle of rotation is strongly chiral-dependent and very large 
 when considering radii a few times larger than the wavelength.
 
Chirality is a geometrical property of structures including  biological molecules \cite{wagniere}, random arrangement of plasmonic  nanostructures \cite{fan2010,pinheiro2017} and nanocrystals \cite{fan2012} that are not superposable with their mirror objects \cite{Wei2017,Moloney}. 
Chirality plays an essential role in several biological, chemical and  nanotechnological applications \cite{Brooks}. Our results for the particle rotation
open the way for the characterization of the local chiral response at the nanoscale,   { that nicely interconnects with already existing local probing techniques to determine viscoelastic properties \cite{Fitzpatrick2018,Lyubin2012} and micro-rheological properties of small particles \cite{Khan2019,power2014,yao2001,En2005}.}

The optical force and torque are strongly dependent on the handedness of the CP trapping laser beam, as expected
since the two different helicities propagate according to different refraction indexes in the chiral host medium. 
Our results indicate that a chiral host medium allows for trapping of high-index particles, and more generally improves the trap axial stability, provided that 
the CP laser beam and the host medium have the same handedness.  

\section*{Results }

 \subsection*{Electromagnetic fields in chiral medium}
 
 The constitutive relations 
 for a chiral medium contain a direct 
 coupling between the electric field $ {\bf  E}$  and the auxiliary field  $ {\bf H} $ proportional to the chirality parameter $ \kappa.$
  They connect the complex displacement field $\textbf{D}$ and magnetic field  $ \textbf{B}$ to  
   $\textbf{E}$ and  $\textbf {H}$ as follows \cite{bohren,chan_nature,lakhtakia,yokota}:

 \begin{gather} 
   \begin{bmatrix} 
 {\bf D}\\  {\bf B} \end{bmatrix}
 = 
  \begin{bmatrix} \epsilon_0  \epsilon & i \sqrt{\epsilon_0 \mu_0}  \, \kappa\\
   -i\sqrt{\epsilon_0 \mu_0}  \,\kappa&\mu_0 \mu 
  \end{bmatrix} \begin{bmatrix}
   {\bf E} \\
   {\bf H}  
   \end{bmatrix}. \label{C7const}
\end{gather} 
where $\epsilon$ and $\mu$ are the relative permittivity and relative permeability of the medium,  respectively. The constant  $\kappa$  is the chirality parameter that characterizes the strength of chirality and usually satisfies the condition $\kappa\ll \sqrt{\epsilon \mu}$. By using the aforementioned constitutive relations the  Maxwell's equations 
for a chiral medium in the frequency domain (frequency $\omega$) can be written in matrix form as
%
 \begin{gather}
  \boldsymbol{\nabla} \cdot\begin{bmatrix}   {\bf E}\\ {\bf H} \end{bmatrix}= 0
\end{gather} 
 \begin{gather}
  \boldsymbol{\nabla} \times  \begin{bmatrix} 
 {\bf E}\\  {\bf H} \end{bmatrix}
 -
 K \begin{bmatrix}
   {\bf E} \\
   {\bf H}  
   \end{bmatrix}=0
\end{gather} 
where \begin{gather} K=  \begin{bmatrix}
   k_0 \kappa & i k_0 \mu \sqrt{\frac{\mu_0}{\epsilon_0}} \\
   -ik_0 \epsilon \sqrt{\frac{\epsilon_0}{\mu_0}}  & k_0 \kappa  
   \end{bmatrix}
   \end{gather}
 and $k_0=\omega/c.$
 The propagation modes are obtained by diagonalizing the matrix $K$ through  a linear transformation \cite{bohren,B1974}
 \begin{gather}
 \begin{bmatrix} {\bf E}\\   {\bf H} \end{bmatrix}
 = \begin{bmatrix}
   1&{-i \mu}/{\epsilon}\\
  {-i \epsilon}/{\mu}&1
   \end{bmatrix}
     \begin{bmatrix}
   {\bf Q}_+ \\
   {\bf Q}_-  
   \end{bmatrix}. \label{C7em in chiral}
\end{gather}
 ${\bf Q}_+$ and ${\bf Q}_-$ independently satisfy the Helmholtz equation
\be 
\nabla ^{2}{\bf Q}_{\sigma}+k_{\sigma} ^{2}{\bf Q}_{\sigma}=0
\ee
and the subsidiary equations
\be
\nabla \times {\bf Q}_{\sigma} + k {\bf Q}_{\sigma}=0 \hspace{10pt}, \hspace{10pt} \nabla \cdot {\bf  Q}_{\sigma}=0 ,
\ee
with $\sigma=\pm 1$ representing helicity. The corresponding wavenumbers
 are given by
\be
k_{\sigma} = k_0 (\sqrt{\epsilon\mu}+\sigma \kappa) .
\ee
Thus, the propagation of a mode of helicity $\sigma$ is governed by the refractive index $n_{\sigma}= \sqrt{\epsilon\mu}+\sigma \kappa.$
%

Finally, for future reference, we define the Debye potentials $\Pi^{\rm E}$ and $\Pi^{\rm M}$ for electric (E) and magnetic (M) multipoles, respectively~\cite{Bowkamp1954,Bobbert1986}:
\begin{equation}
 \begin{aligned}
{\bf E}=  \nabla \times \nabla \times\left({\bf r} \Pi^{\rm E}\right)+i\omega \mu\mu_0 \nabla \times\left({\bf r} \Pi^{\rm M}_{}\right),\\
{\bf H}=  \nabla \times \nabla \times\left( { \bf r}\Pi^{\rm M}  \right)-i\omega \epsilon\epsilon_0  \nabla \times \left( {\bf r}\Pi^{\rm E}_{}\right).
\end{aligned} \label{C7Field_tot}
\end{equation}

\subsection*{The electromagnetic stress tensor in a chiral medium}

When a light beam is scattered off a particle, it imparts an optical force on it due to momentum conservation.  Such force may be calculated by integrating the Maxwell stress tensor $\overset{\leftrightarrow}{\bf T}$ over a closed Gaussian surface $S$ that wraps around the particle:
\begin{equation}
{\bf F}=   \oint_S  \langle \overset{\leftrightarrow}{\bf T} \rangle \cdot \mathbf{dA}.
\label{C7force}
\end{equation} 
As the stress tensor is quadratic in the electromagnetic fields, the following identity \cite{Jackson} is useful when evaluating
the time average $\langle ... \rangle$ in  (\ref{C7force}):
\begin{equation}
\langle \pmb{\cal{V}}_1({\bf r},t) \, \pmb{\cal{V}}_2({\bf r},t) \rangle = \frac{1}{2} \,{\rm Re} \, \left[  {\bf V}_1 ({\bf r}) \, {\bf V}_2^*({\bf r}) \right]
\label{average}
\end{equation} 
where $\pmb{\cal{V}}_j= {\rm Re}({\bf V}_j\,e^{-i\omega t})$ $j=1,2$
are general monocromatic fields. 

We evaluate the surface integral in (\ref{C7force}) for a Gaussian spherical surface $S(R)$ of radius $R$ centered at the origin. 
Using (\ref{average}) and taking
the standard explicit expression for the stress tensor $ \overset{\leftrightarrow}{\bf T}$ in a non-viscous, incompressible liquid at rest \cite{Robinson,Pfeifer07}, 
we find \footnote{{There is a disagreement in the Minkowski and Abraham prescriptions, but it manifests itself in the momentum {\it density}, not in the momentum {\it flux}, so the stress tensor is the same.}} 
\begin{eqnarray}\label{Total_force}
{\bf F}&=&\frac{R^2}{4}\oint_{S(R)}d\Omega\; {\rm Re} \Bigl[\textbf{E}\left(\textbf{D}^{*}\!\! \cdot \hat{\bf r}\right) +  \textbf{D}\left(\textbf{E}^{*}\!\! \cdot \hat{\bf r}\right)
 +\,  \textbf{H}\left(\textbf{B}^{*}\!\! \cdot \hat{\bf r}\right) +  \textbf{B}\left(\textbf{H}^{*}\!\! \cdot \hat{\bf r}\right) 
- (\textbf{E} \cdot \textbf{D}^{*}+\textbf{H} \cdot \textbf{B}^{*}) \hat{\bf r} \Bigr]. 
\end{eqnarray}
Finally, replacing the constitutive relations (\ref{C7const}) for a non-magnetic chiral medium
into (\ref{Total_force})
 and taking $R \rightarrow \infty$, we arrive at
\begin{eqnarray}
 {\bf F} = -\lim_{R\rightarrow \infty}\frac{R^2}{2} \int_{S(R)} \!\!\! d\Omega \, \hat{\bf r} \biggl[\frac{\epsilon_0\epsilon {\bf E\cdot E^{*}} + \mu_0 {\bf H\cdot H^{*}}}{2} 
- \sqrt{\epsilon_{0}\mu_{0}}\, \kappa \, {\rm Im}(\mathbf{E}^{*}\cdot\mathbf{H}) \biggr] , 
\label{C7max}
 \end{eqnarray}
where we have used that the radial field components decay as $1/r^2$ and hence do not contribute to the flux.

\subsection*{Theory of optical tweezers of a dielectric sphere embedded in a chiral medium}

Here we combine the previous results in order to
develop a generalization of the Mie-Debye theory to the case of a chiral host medium.
We consider a CP Gaussian laser beam 
 of helicity $\sigma=\pm 1$ at the entrance port of a high-numerical aperture (NA)
 objective.
 The resulting non-paraxial focused beam 
 propagates in the lossless non-magnetic chiral medium 
 of refractive index $n_{\rm m}(\sigma)=\sqrt{\epsilon_{\rm m}}+\sigma \kappa$, 
 where $\epsilon_{\rm m}$
is the relative permittivity and 
$\kappa$ is the chirality parameter. 

The focused beam is then
 represented as a superposition of plane waves corresponding to wavevectors ${\bf k}_{\sigma}(\theta,\phi)$ with a fixed magnitude $k_{\sigma}=n_{\rm m}({\sigma})k_0$~\cite{RichardsWolf1959}:
\begin{eqnarray}
\mathbf{E}_{\rm in}^{(\sigma)}(\mathbf{r})&=&{E}_{0}\int_{0}^{2\pi}d\phi\int_{0}^{\theta_{0}}d{\theta}\sin \theta\, \sqrt{\cos \theta}\,e^{-\gamma^2\sin^2\theta}
 \, e^{i{\bf k}_{\sigma}(\theta,\phi)\cdot(\mathbf{r}+\mathbf{r}_{p})}\,\hat{\boldsymbol{\epsilon}}_{\sigma}^{\prime}(\theta, \phi).\label{C7campofocalizado}
\end{eqnarray}
The polarization unit vector  $ \hat{\boldsymbol{\epsilon}}_{\sigma}^{\prime}(\theta, \phi) =(\hat{\bf x}'+
 i\,\sigma\, \hat{\bf y }')/\sqrt{2}$  
is defined  in terms of the 
Cartesian unit vectors obtained 
by  rotation with Euler angles
 $ (\phi,\theta,-\phi)$.
 The focal point is at position $-\mathbf{r}_{p}$, whereas the spherical particle center is at the origin~\cite{Mazolli2003}. After resolving for the Mie scattering by the particle
  and computing the resulting optical force, we displace the origin to the focal point, and then the particle position will be at 
  ${\bf r}_p(\rho_p,\phi_p,z_p)$ which is finally written in terms of its cylindrical components. 
  The angular semi-aperture $\theta_0$ is defined  in terms of the objective NA as discussed in detail below, whereas 
 $\gamma$ is the ratio of the objective focal length to the laser beam waist at the objective entrance port. 
   
   The incident focused beam (\ref{C7campofocalizado}) illuminates 
an achiral, non-magnetic spherical particle of refractive index $n_p=\sqrt{\epsilon_p}$ and radius $a$ which is embedded in the chiral medium. 
Mie scattering of a plane wave by a chiral spherical particle embedded in a chiral medium has been solved in Ref.~\cite{Hinders1992}.
We consider the particular case in which only the host medium is chiral.
Our results differ from \cite{Park2015} by some sign factors which
we attribute to typos in that reference. 

When considering the non-paraxial focused beam (\ref{C7campofocalizado}), 
we need to solve the Mie scattering for different propagation directions and then take the superposition of 
the corresponding scattered field components. Due to the spherical symmetry of the particle, 
it is straightforward to write the corresponding scattering field components with the help of finite rotations and Wigner rotation matrix elements
  $d_{m,m'}^{\ell}(\theta)$ in the angular momentum representation~\cite{Edmonds}.
  
 The Debye potentials describing the total fields outside the spherical particle 
 are written as partial-wave (multipole) sums over $\ell$ (for the total angular momentum $J^2$) and $m$ (for the axial component $J_z$) of the form
 \[
\sum_{\ell,m}\equiv  \sum^{\infty}_{\ell=1}\, \sum^{\ell}_{m=-\ell}.
 \]
Denoting the spherical coordinates of a spatial point ${\bf r}$ as $(r,\vartheta,\varphi),$ we find
\begin{eqnarray}
\Pi^{\rm E}(r,\vartheta,\varphi) & =  i\sigma \frac{E_0}{k_\sigma}& \sum_{\ell m} \gamma^{(\sigma)}_{\ell,m}\bigg[\, j_\ell(k_\sigma r )
+  A_{\ell}\,  h^{(1)}_\ell(k_\sigma r ) - \, B_{\ell} \,  h^{(1)}_\ell(k_{-\sigma} r )\bigg]Y_{\ell,m}(\vartheta ,\varphi) 
\label{PiEexplicit}
 \end{eqnarray}
\begin{eqnarray}
\Pi^{\rm M}(r,\vartheta,\varphi) & =   \sqrt{\frac{\epsilon_m\epsilon_0}{\mu_0}}\frac{E_0}{k_\sigma} &  \sum_{\ell m}\gamma^{(\sigma)}_{\ell,m} \bigg[\,  j_\ell(k_\sigma r ) 
 + A_{\ell}\,  h^{(1)}_\ell(k_\sigma r ) 
 +\, B_{\ell}\,  h^{(1)}_\ell(k_{-\sigma} r )\bigg]Y_{\ell,m}(\vartheta ,\varphi)   \label{PiMexplicit}
 \end{eqnarray}
 Here, $ j_{\ell} $ and $ h^{(1)}_{\ell}$ denote the spherical Bessel and Hankel functions  of the first kind, respectively, and 
 $Y_{\ell,m}$ are the spherical harmonics~\cite{DLMF25.12}. The multipole coefficients of the incident focused beam  (\ref{C7campofocalizado})
 are given by
\begin{eqnarray}\label{gamma}
\gamma^{(\sigma)}_{\ell,m}&=& 2\pi\, i^{\ell-m+\sigma}\sqrt{\frac{4\pi(2{\ell}+1)}{{\ell}({\ell}+1)}}e^{-i(m-\sigma)\phi_p} 
\int_{0}^{\theta_{0}}d\theta\sin\theta\sqrt{\cos\theta}d^{\ell}_{m,\sigma}(\theta)J_{m-\sigma}(k_\sigma\rho_p\sin\theta_{})
e^{ik_\sigma z_p \cos\theta}
\end{eqnarray}
where $J_{m}$ are the cylindrical Bessel functions of integer order $m$ \cite{DLMF25.12}. 
The scattering Mie coefficients $A_{\ell}\equiv\alpha_{\ell}/\Delta_{\ell}$ and 
 $B_{\ell}\equiv\beta_{\ell}/\Delta_{\ell}$ 
 represent the amplitudes for conserving and changing 
 the photon helicity, respectively. 
 They are given by 
\begin{eqnarray}\nonumber
\alpha_{\ell}&=&(N^{2}+1)\,
\left[\psi_\ell(x)\xi^{\prime}_\ell(\bar{x})+ \xi_\ell(\bar{x}) \psi^{\prime}_\ell(x) \right]\psi_\ell(y)\psi^{\prime}_\ell(y)
-2N\left[\psi^{\prime}_\ell(x) \xi^{\prime}_\ell(\bar{x})\psi_\ell(y)^{2}+\psi_\ell(x) \xi_\ell(\bar{x})\psi^{\prime}_\ell(y)^2 \right]\nonumber\\
\\
\nonumber
\beta_{\ell}&=& (N^{2}-1) \left[ \psi_\ell(x)\xi^{\prime}_\ell(x) -\xi_\ell(x)\psi^{\prime}_\ell(x)\right] \psi_\ell(y)\psi^{\prime}_\ell(y)\\
\nonumber\\
\nonumber
\Delta_{\ell}&= &\left[(N^{2}+1)\xi^{\prime}_\ell(x)\psi_\ell(y) - 2N\xi^{}_\ell(x)\psi^{\prime}_\ell(y)\right]\xi_\ell(\bar{x})\psi^{\prime}_\ell(y)
  +\left[(N^{2}+1) \xi_\ell(x)\psi^{\prime}_\ell(y) -2N \xi^{\prime}_\ell(x) \psi_\ell(y)\right] \xi^{\prime}_\ell(\bar{x}) \psi_\ell(y) 
\end{eqnarray}
Here,  $N=n_p/n_{\rm m}(\sigma)$  is the relative refractive index of the particle with respect to the host medium { and
 $\psi_\ell(x)=xj_\ell(x)$ and $\xi_\ell=xh^{(1)}_\ell(x)$ are the Riccati-Bessel functions \cite{bohren}.
 The variables 
 $x=n_{\rm m}(\sigma)  k_0 a$ and $\bar{x}=n_{\rm m}(-\sigma)  k_0 a$ are the size parameters in the host medium
when taking the incident and the reversed helicities, respectively. 
 Finally, $y=N\,x$ is
  the size parameter in the particle medium.}
 
 The coefficients $B_{\ell}$
 appearing  in Eqs.~(\ref{PiEexplicit}) and (\ref{PiMexplicit}) 
  represent amplitudes for 
 helicity reversal $\sigma\rightarrow -\sigma$ upon Mie scattering.
   When the microsphere is aligned along the symmetry $z-$ axis ($\rho_p=0$), the total optical angular momentum is conserved, 
 and as a consequence the variation of SAM is entirely converted into optical orbital angular momentum \cite{Schwartz2006}. 
Mie scattering is indeed a mechanism for spin-orbit interaction~\cite{Haefner2009,Bliokh2015}.
On the other hand, when $\rho_p>0,$ part of the SAM change might be transferred to the particle center-of-mass, thus contributing to the 
optical torque on the particle. 

 We now define the normalised force efficiency~\cite{ashkin1992}
 \begin{equation}
 \label{normalization}
 {\bf Q}= {\bf F}/(n_{\rm m}(\sigma) P/c)
 \end{equation}
  where 
  $P$ is the laser power in the sample region and $c$ is the speed of light in vacuum. 
  When evaluating the flux of the stress tensor (\ref{C7max}), we write the total electric field as 
\begin{equation}
\textbf{E}= \textbf{E}^{(\sigma)}_{\rm in}+\textbf{E}^{(\sigma)}_{\rm s}+\textbf{E}^{(-\sigma)}_{\rm s} \label{C7et}
\end{equation}
  and likewise for the magnetic field ${\bf H}.$
 $\textbf{E}^{(\sigma)}_{\rm s}$ ($\textbf{E}^{(-\sigma)}_{\rm s}$) represents the scattered field contribution with the same (opposite) helicity of the incident field.
 The three terms in (\ref{C7et}) are ordered precisely as the three contributions in the r.-h.-s. of (\ref{PiEexplicit}) and (\ref{PiMexplicit}).
 
 Among the several quadratic contributions obtained when replacing (\ref{C7et}) into 
 (\ref{C7max}), cross terms involving opposite helicities do not contribute and only  $ \textbf{E}^{(\sigma)}_{\rm in}\cdot\textbf{E}^{(\sigma)}_{\rm s}{}^*,$   $\textbf{E}^{(\sigma)}_{\rm s}\cdot\textbf{E}^{(\sigma)}_{\rm s}{}^*$ and $\textbf{E}^{(-\sigma)}_{\rm s}\cdot\textbf{E}^{(-\sigma)}_{\rm s}{}^*$
  survive when taking
 a Gaussian surface at infinity (and likewise for  the terms quadratic in ${\bf H}$ and the cross electric-magnetic terms). 
 The first
 term yields the extinction contribution ${\bf Q}_{\rm e}$ to the force efficiency,
 while the last two terms yield the scattering contribution ${\bf Q}_{\rm s}.$
 ${\bf Q}_{\rm e}$ represents
 the rate at which linear momentum is removed from the incident field,  normalized as in (\ref{normalization}). 
 A fraction of this momentum is carried away by the scattered field at a normalized rate $-{\bf Q}_{\rm s},$
 so that the total force efficiency is written as 
 \begin{equation}
{\bf Q}={\bf Q}_{\rm e}+{\bf Q}_{\rm s}.
\end{equation}
 
 When deriving the multipole series for ${\bf Q}_{\rm e}$ and ${\bf Q}_{\rm s}$  from
 Eqs.~(\ref{C7max}) and (\ref{PiEexplicit})-(\ref{gamma}),
 we introduce the effect of refraction at the planar interface between the glass coverslip and the chiral medium filling the sample, which is typical in oil-immersion objectives \cite{Viana2007}.
 The refraction index mismatch between the two media is written in terms of the relative index
  $N_\sigma=n_{\rm m}(\sigma)/n_{\rm g},$
  where $n_{\rm g}$ is the glass refractive index. 
  The Fresnel amplitude for refraction is 
    \[
  T(\theta)= \frac{2\cos\theta}{\cos\theta+N_\sigma \cos \theta_{\rm m}}
  \]
  The wavevectors in glass have magnitude $k_{\rm g}=n_{\rm g}k_0$
  and make an angle $\theta$ with respect to the $z$-axis, whereas the angle in the 
  chiral medium  is 
 $\theta_{\rm m}=\arcsin(\sin\theta/N_\sigma).$
  The laser power $P$ in the sample region is reduced 
on account of the interface, as well as from the finite aperture radius of the objective entrance port. The resulting filling fraction
 is given by
\begin{equation*}
F_{\sigma} = 16\gamma^2\int_0^{s_0} ds\,s\,e^{(-2\gamma^2s^2)}\,\frac{\sqrt{(1-s^2)(N_{\sigma}^2-s^2)}}{\left(\sqrt{1-s^2}+\sqrt{N_{\sigma}^2-s^2}\right)^2},
\end{equation*}
 with $s_0=\min\{N_{\sigma},\mbox{NA}/n_{\rm g}\}$. 
 
 More importantly, 
 we add to the ideal aplanatic model 
(\ref{C7campofocalizado}) the spherical aberration phase~\cite{Torok1995}
\begin{equation}
\Phi_{\rm sa}(\theta)=k_{\rm g}\left( -L/N_{\sigma}\cos\theta+N_{\sigma}L\cos\theta_{\rm m}\right), \label{C7aberration_interface}
\end{equation}
also introduced by refraction at the glass-sample interface. Spherical aberration is typically detrimental on the effects discussed in this paper as it degrades the focal region.
Thus, our realistic description of oil-immersion objectives, which are usually employed in optical tweezers setups, prevents us from overestimating the optical torque discussed in the following. 
The phase $\Phi_{\rm sa}(\theta)$ is proportional to the distance $L$
 between the glass slide and  paraxial focal plane. Instead of assuming a given value for $L,$ which is unknown in real experiments, we simulate the experimental procedure 
 for controlling the amount of spherical aberration (see Methods). The relevant length scale is then the distance $d$ by which the objective is displaced, starting from
the configuration with the trapped microsphere just touching the coverslip. 

The axial components of  ${\bf Q}_{\rm s}$ and ${\bf Q}_{\rm e}$  are then written as

%
%
%
\begin{eqnarray}\nonumber
Q_{s z} &=& -\frac{16\gamma^2}{F_{\sigma}N_{\sigma}}{\rm Re}\Biggl\{\sum_{\ell m}\frac{\sqrt{\ell(\ell+2)(\ell+m+1)(\ell-m+1)}}{\ell+1} 
  \biggl[\biggl(1+\frac{\sigma\,\kappa}{n_{\rm m}}\biggr)A_{\ell}A_{\ell+1}^{*}+\biggl(1-\frac{\sigma\,\kappa}{n_{\rm m}}\biggr)B_{\ell}B_{\ell+1}^{*}\biggr] \\
&& \times  G^{(\sigma)}_{\ell,m}G^{(\sigma)*}_{\ell+1,m} 
+\frac{1}{2}
\frac{(2\ell+1)}{\ell(\ell+1)}\,m\,\sigma\,  \biggl[\biggl(1-\frac{\sigma\,\kappa}{n_{\rm m}}\biggr)\vert A_{\ell}\vert^2-\biggl(1+\frac{\sigma\,\kappa}{n_{\rm m}}\biggr)\vert B_{\ell}\vert^2\biggr]\vert G^{(\sigma)}_{\ell,m}\vert^2\Biggr\}, \label{C7Qszp}
\\
Q_{ez}&=&\frac{8\gamma^2}{F_{\sigma}N_{\sigma}}{\rm Re}\sum_{\ell m}(2\ell+1)
\biggl(1-\sigma\frac{(i-1)}{2}\frac{\kappa}{n_{\rm m}}\biggr)\,A_{\ell}\,G^{(\sigma)}_{\ell,m}\,G'^{(\sigma)*}_{\ell,m},
\label{C7Qez}
\end{eqnarray}

Like the axial components,
the radial and azimuthal cylindrical components of ${\bf Q}_{\rm s}$ and ${\bf Q}_{\rm e}$
do not depend on the 
particle angular position $\phi_p$ by symmetry (see Methods). 
The dependence on the cylindrical coordinates $\rho_p$ and $z_p$ of the particle position is contained in the multipole coefficients 
\begin{eqnarray}\label{C7Gjm}
G_{\ell m}^{(\sigma)}(\rho_p,z_p)=\int_{0}^{\theta_0}d\theta\sin\theta\sqrt{\cos \theta}\,T(\theta)\, e^{-\gamma^2\sin^2\theta}
d_{m,\sigma}^{\ell}(\theta_{\rm m})\, J_{m-\sigma}\left( k_{\rm g} \rho_p\sin\theta\right) 
e^{i[\Phi_{\rm sa}(\theta)+k_{\sigma}\cos\theta_{\rm m} z_p ]}\\
G'^{(\sigma)}_{\ell,m}(\rho_p,z_p)=\int_{0}^{\theta_0}d\theta\sin\theta \cos\theta_{\rm m}  \sqrt{\cos \theta}\,T(\theta)\,e^{-\gamma^2\sin^2\theta}d_{m,\sigma}^{\ell}(\theta_{\rm m})\, J_{m-\sigma}\left( k_{\rm g}\rho_p \sin\theta\right)  e^{i[\Phi_{\rm sa}(\theta)+k_{\sigma}\cos\theta_{\rm m} z_p ]}\label{C7multipole coefficient}
\end{eqnarray}


The upper bound for the integration  in (\ref{C7Gjm}) and (\ref{C7multipole coefficient}) represents the angular semi-aperture in the glass medium:
$\theta_0=\sin^{-1}\{{\rm min}[({\rm NA}/n_{\rm g}),N_\sigma]\}.$
 We do not take evanescent waves
 (which appear when ${\rm NA}> n_{\rm m}(\sigma)$)
  into account~\cite{Dutra2016} as the microsphere is trapped near the focal plane and thus far from the coverslip for 
  the typical numerical examples discussed below.

\subsection*{Numerical examples}

In all numerical examples discussed in this paper,
we take typical experimental values for a standard optical tweezers setup \cite{Dutra2014}.
 We consider a right-handed CP Gaussian beam 
 ($\sigma=-1$)
  with vacuum wavelength  $\lambda_0=1064\,{\rm nm}$ at the objective entrance port, 
 of numerical aperture NA$=1.3.$ 
 Results for left-handed CP  ($\sigma=1$) may be considered from those shown here by replacing $\kappa\rightarrow -\kappa$ and  
 changing the sign of the azimuthal force component (and its derivative). 
 The ratio between the objective focal length and the beam waist at the entrance port is $\gamma=1.226.$
The objective axial displacement, which controls the amount of spherical aberration when employing oil-immersion objectives, is 
 $d=5\,\mu{\rm m}.$
 In addition, we take $ n_{\rm g}=1.5$ and $\epsilon_{\rm m}= 1.85$ for 
 the refractive index of the glass coverslip 
 and the permittivity of the chiral solution, respectively.

\begin{figure}
\centering
\includegraphics[width =5.in]{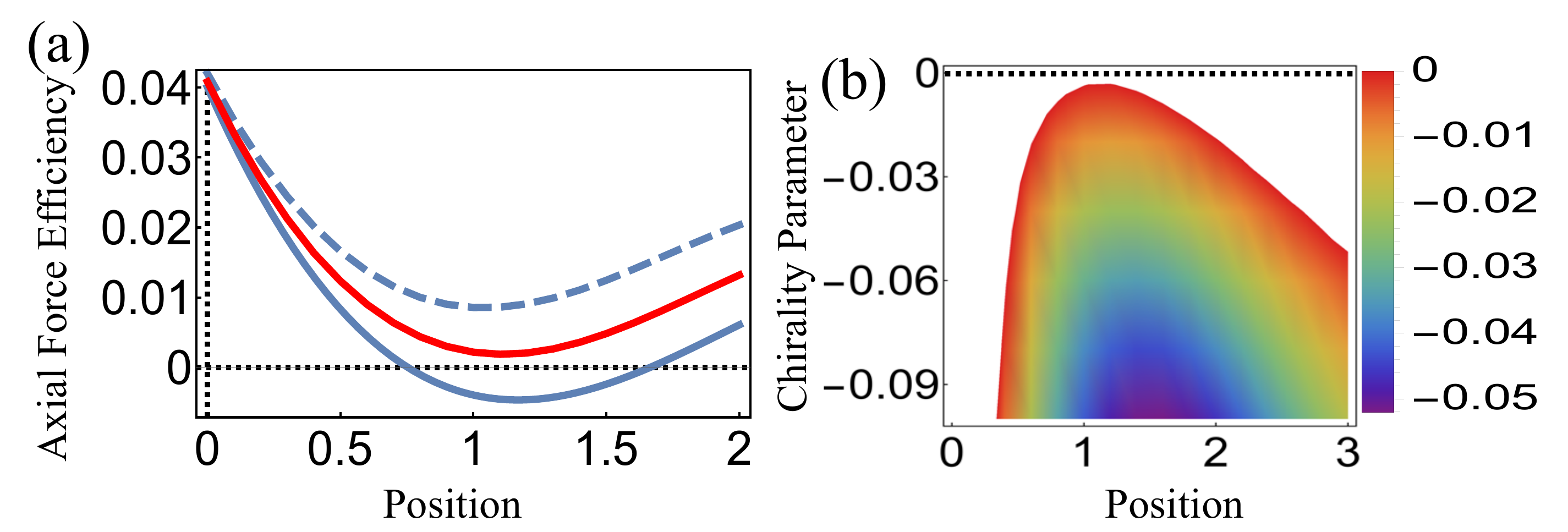}
\caption{ (a) Normalized axial force $Q_z$ acting on a BrO$_2$ microsphere of radius $500\,{\rm nm}$ embedded in a chiral medium as  a function of axial position (in units of the sphere radius)  for different chirality parameters: $ \kappa = -0.01$ (solid), $\kappa$ = 0 (red)  and  $\kappa = 0.01 $ (dashed).  The incident beam is right-handed circularly polarized (helicity $\sigma=-1$).
(b)  Density plot of $Q_z$ 
versus axial position and chirality parameter. Only negative values are shown.} 
 \label{C7F2}
\end{figure}

\textbf{Optical Force.}
As a first example, we consider a BrO$_2$ microsphere with refractive index $n_p=1.7$ and radius $a = 500\, {\rm nm}$ 
immersed in a chiral solution.  
In  Fig. \ref{C7F2}(a),  we plot the axial force efficiency $Q_z$ as a function of normalized microsphere position $z_p/a$
along the $z-$ axis ($\rho_p=0$)
 for different values of chirality parameter $\kappa$.
   When the host medium is achiral ($\kappa=0$),
  its refractive index  $n_{\rm m}=\sqrt{\epsilon_{\rm m}}=1.36$ is too small compared to the { BrO$_2$ particle's high refractive index.} As a consequence, radiation pressure 
   dominates, leading to a positive (i.e. along the propagation direction) force for all values of $z_P/a$ (red line). 
   
In contrast, trapping can be achieved in a chiral host media
with $\kappa=-0.01$
 under otherwise the same conditions, 
as indicated by the solid blue line in Fig.~\ref{C7F2}(a).   Thus, a chiral medium with the same handedness of 
the CP trapping laser beam allows for trapping of { high-index particles} by diminishing the radiation pressure effect and leading to negative optical forces. On the other hand, 
in the case of a left-handed chiral medium $\kappa_m=0.01$  (dashed  line), 
radiation pressure is enhanced and again no trapping is possible. 

 Fig. \ref{C7F2}(b)  shows the density plot of the
 axial force efficiency
 $Q_z$ as a function of the chirality parameter $\kappa$ and the axial position $z_p/a.$ The colored area corresponds to
 the regions in the parameter space for which the optical force is negative ($Q_z<0$), thus allowing for stable trapping.  
 The edge of this area provides the positions of equilibria along the $z$-axis as function of $\kappa,$ with the left-hand side corresponding to stable equilibria. 
 It is worthwhile to mention that chiral media not only optimize trapping stability but also facilitate optical tweezing of large refractive-index particles, as in the example considered here. In short, trapping in a chiral host media facilitates optical manipulation and tweezing of  high-index particles provided that the chiral material has the same handedness of the incident CP light.

\textbf{Optical Torque.}
Spin angular momentum (SAM) of CP light can be transferred to trapped particles and make them spin around the beam axis when 
they are absorptive, anisotropic \cite{Friese1998} or non-spherical \cite{Bishop2003}. Although the optical torque (OT) on a transparent isotropic microsphere centered along the beam symmetry axis vanishes, transfer of SAM to the center of mass can still be observed in this case from the analysis of Brownian fluctuations \cite{Ruffner2012} or by  driving the sample laterally so as to displace the equilibrium position from the beam axis \cite{Diniz2019}. 
OT is predicted to be significantly enhanced in the case of chiral particles, opening the way for enantioselective manipulation and characterization
of the chiral response of individual nanoparticles 
 with optical tweezers~\cite{rali2020}. 

\begin{figure}
\centering
\includegraphics[width = 3.4in]{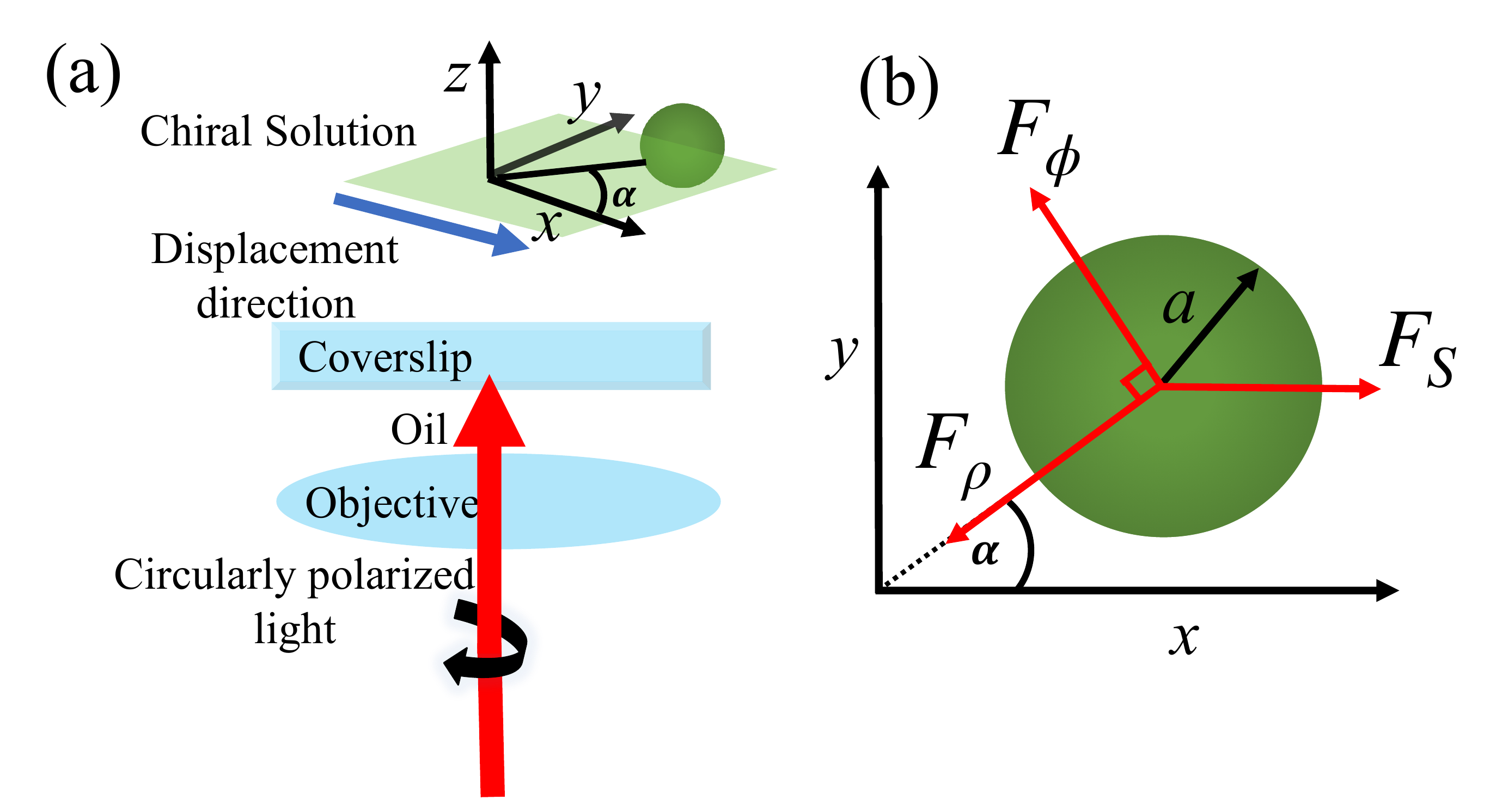}
\caption{ (a)   Schematic representation of the optical torque on a particle trapped in a chiral medium. A right-handed ($\sigma=-1$) 
circularly-polarized Gaussian laser beam is focused  by an oil-immersion high-NA objective into a sample filled with a chiral solution.
The sample is driven laterally so as to displace the particle equilibrium position from the beam symmetry axis.
  (b) At equilibrium, the resulting  Stokes drag force $F_S$ balances the optical force, which contains radial 
  $F_{\rho}$
  and azimuthal 
  $F_{\phi}$ components, the latter being responsible for the optical torque. The equilibrium position is then rotated around the beam axis by an angle $\alpha$ with respect 
  to the direction of the Stokes force.} 
 \label{C7F1}
\end{figure}

Here we show that a much stronger enhancement of the OT is found when taking a chiral host medium instead of a chiral particle.
We follow the scheme of Ref.~\cite{Diniz2019} and calculate the rotation of the equilibrium position when a Stokes drag force is applied  by driving the sample 
along the $x-$ direction, as illustrated by Fig.~\ref{C7F1}(a). 
As the particle is displaced off-axis by the Stokes force $F_S$, an optical  azimuthal component $F_{\phi}$ builds up, 
in addition to the restoring radial component $F_{\rho}<0.$ 
$F_\phi$ results from the SAM of the trapping beam and its sign is controlled by the helicity $\sigma$ of the CP.
As shown in Fig.~\ref{C7F1}(b), 
the resulting equilibrium position is then rotated by an angle $\alpha$ around the $z-$axis with respect to the 
$x-$axis, with $\tan\alpha = F_\phi/|F_\rho|.$
When the off-axis displacement is $\ll a,$ we can write the rotation angle in terms of the transverse optical stiffness $k_{\rho}\equiv -\partial_\rho F_\rho|_{\rho=0}$ and the 
torsion constant $k_{\phi}\equiv\partial_\rho F_\phi|_{\rho=0}$ as 
 $\tan\alpha \approx k_{\phi}/k_{\rho}.$
 We obtain exact  values for $k_\phi$ and $k_\rho$ from the Mie-Debye theory for chiral host media developed above. 
We first derive partial-wave series for $k_\phi$ and $k_\rho$ by taking the analytical derivatives of 
the series for $Q_\phi$ and $Q_\rho.$ The series for $k_\phi$ and $k_\rho$ are then computed numerically. 
 By rotational symmetry, they are independent of the angular position $\alpha.$ 
 
\begin{figure}
 \centering
\includegraphics[width = 3.in]{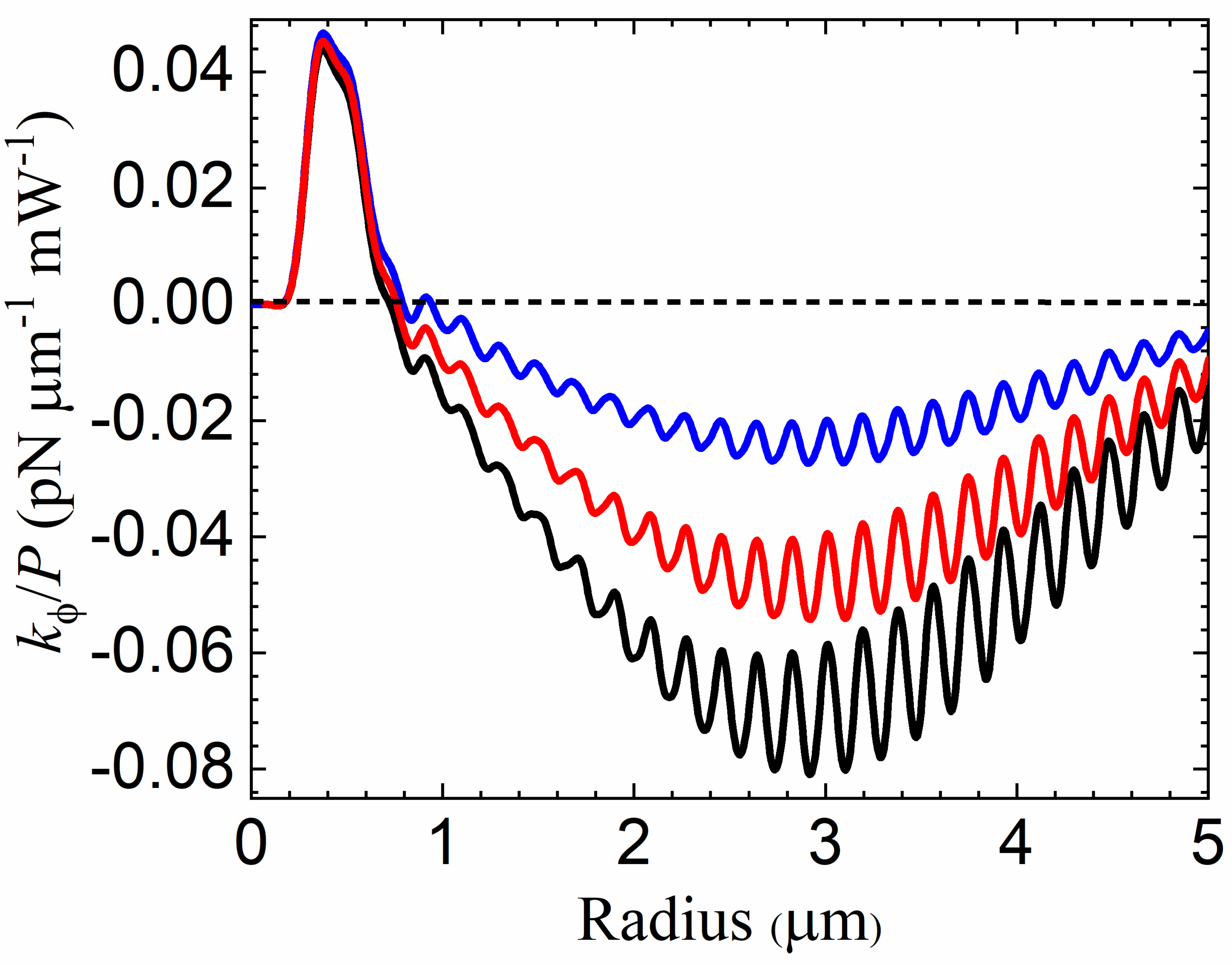}
\caption{Torsion constant per unit power $k_{\phi}/P$ 
as a function of the radius of a silica microsphere.
The particle is embedded in a chiral solution with
 $\kappa=-0.001$ (blue), $-0.002$ (red) and $-0.003$ (black).
 The incident beam is right-handed circularly-polarized (helicity $\sigma=-1$). 
 }  \label{kphi}
\end{figure}

In all numerical examples for the optical torque, we consider a silica microsphere with refractive index  $n_p = 1.46$ embedded in a chiral medium.
In Fig.~\ref{kphi}, we plot $ k_{\phi}/P$ as a function of the sphere radius $a$ 
for chirality parameters $\kappa=-0.001$ (blue), $-0.002$ (red) and $-0.003$ (black).
For radii $a\stackrel{<}{\scriptscriptstyle\sim} \lambda_0/n_m(\sigma),$ 
the torsion constant is approximately independent of $\kappa$ and develops a peak corresponding to a negative torque \cite{Magallanes2018}, i.e, opposite to the SAM of the 
trapping beam, at $a\approx 0.4\,\mu{\rm m}.$  
As the radius increases, the OT goes positive and $k_\phi$ develops a second (negative) peak 
at $a\approx 2.9\,\mu{\rm m}$ whose amplitude is strongly chirality-dependent.   
For even larger radii (not shown), $k_\phi$ oscillates around zero as expected in the geometrical optics regime~\cite{Mazolli2003}. 
The oscillations result from interference between direct reflection and reflection after a round-trip propagation across the microsphere diameter~\cite{Neto2000}.
Such interference oscillations, of 
 period $\Delta a= \lambda_0/(4n_p)\approx 0.18\,\mu{\rm m},$  are clearly visible in the negative peak shown in 
 Fig.~\ref{kphi}.

The spin-orbit contribution to $k_\phi$ can be traced by collecting the spin-reversal terms involving the coefficients $B_{\ell}$
 as discussed in connection with Eqs.~(\ref{PiEexplicit}) and (\ref{PiMexplicit}). In the peak around
$a\approx 0.4\,\mu{\rm m}$ shown in Fig.~\ref{kphi}, the spin-orbit contribution is negative and its magnitude varies in the range between $15\%-20\%$ 
of the total result. It becomes more dominant for smaller particles, closer to the Rayleigh scattering regime, for which $k_\phi$ becomes negligibly small and  $\kappa-$independent. 
On the other hand, the spin-orbit effect accounts for a small fraction of the torsion constant $k_\phi,$ typically at the percent level, near the chirality-dependent peak 
around  $a\approx 2.9\,\mu{\rm m}.$ Overall, the relative contribution of the spin-orbit term tends to decrease with the chirality parameter $\kappa.$

\begin{figure}
\centering
\includegraphics[width = 3.in]{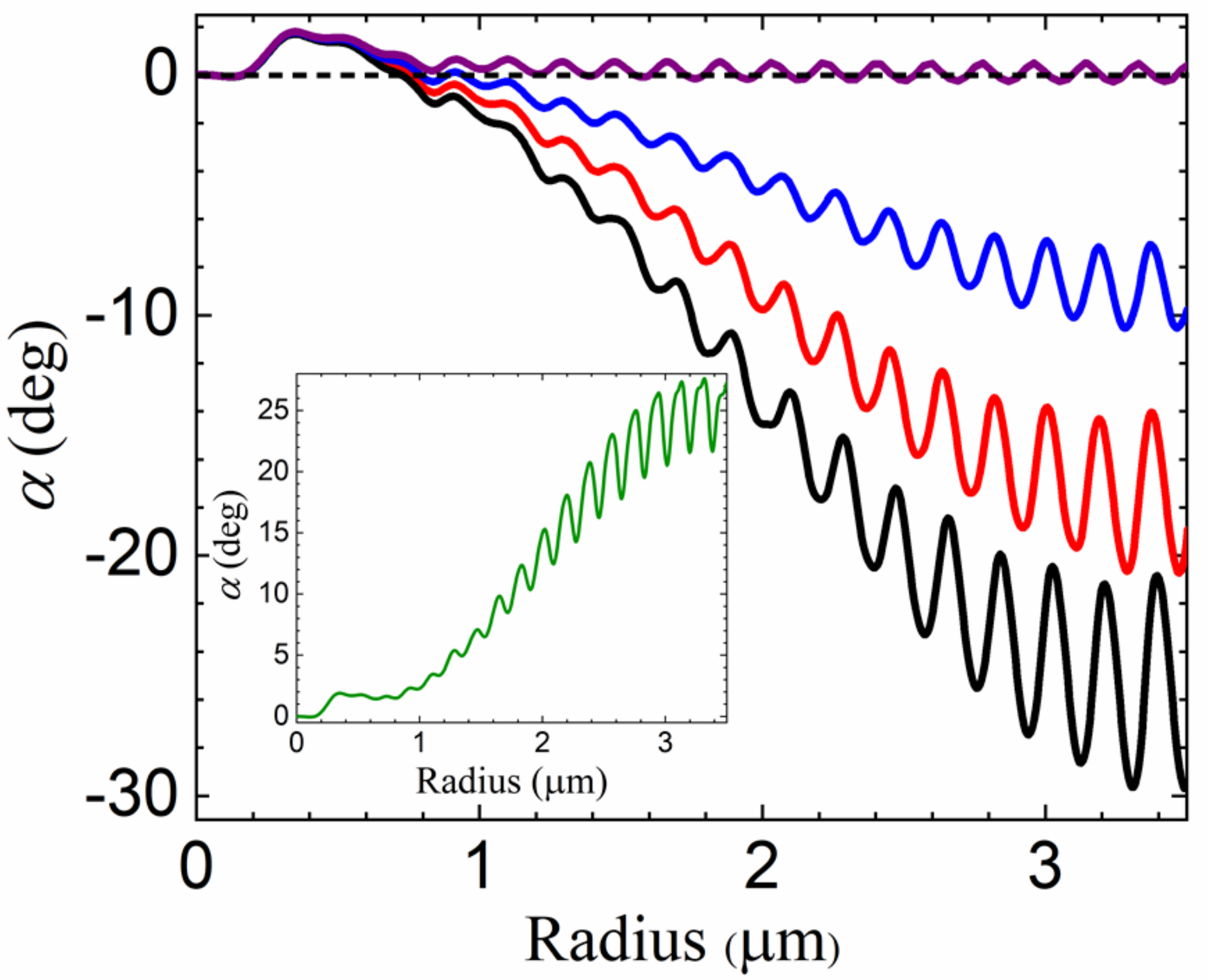}
\caption{ 
Microsphere rotation angle $\alpha$ in degrees resulting from the optical torque on a silica microsphere (see Fig.~\ref{C7F1}) as a function of radius. 
The chirality parameter of the host medium is
$\kappa=0$  (purple),
 $-0.001$ (blue), $-0.002$ (red) and $-0.003$ (black). The inset shows the case of a left-handed 
host medium with  $\kappa=0.003$ (green) for comparison. The incident beam is right-handed circularly-polarized (helicity $\sigma=-1$).  The incident beam is right-handed circularly-polarized (helicity $\sigma=-1$). }  \label{C7F3}
\end{figure}

Such chirality-dependent enhancement of the OT illustrated by Fig.~\ref{kphi} leads to a significant increase of the rotation angle $\alpha.$
In Fig. \ref{C7F3}, we plot $\alpha$  as a function of sphere radius, again for different values of the chirality parameter (same conventions as in Fig.~\ref{kphi}).   
We also show the case of an achiral medium ($\kappa=0$, purple), for which
 the rotation is significant only for radii near $a\sim 0.4\,\mu{\rm m},$ resulting from a negative OT recently measured for polystyrene microspheres \cite{Diniz2019}. 
For chiral media, the magnitude of the rotation angle is 
 maximum at $a\approx 3.4\,\mu{\rm m},$ 
which is slightly shifted with respect to the peak position of $k_{\phi}$ because the transverse stiffness decays as $k_\rho \sim 1/a$ in this size range \cite{Mazolli2003}.
The behavior of $k_\rho$ also explains why the ratio between the amplitudes of the two peaks for the angle of rotation
is much bigger than the corresponding ratio for the torsion constant $k_\phi$
 shown in Fig.~\ref{kphi}.

The interference oscillations discussed in connection with Fig.~\ref{kphi} are also clearly visible in the plot of the rotation angle shown in Fig.~\ref{C7F3}.
The fast, large-amplitude oscillations near the peak region open the way for measurements of the microsphere radius.  
For $\kappa = -0.003,$ 
the oscillations 
correspond to a maximum slope 
$\Delta \alpha/\Delta a \sim 130^{\rm o}\mu{\rm m}^{-1}$
near the peak region,
 allowing for a sensitivity $\delta a\sim 1.5\,{\rm nm}$ given a typical conservative estimate
$\delta \alpha \sim 0.2^{\rm o}$ for the precision in the measurement of the rotation angle \cite{Diniz2019}. 

The inset of Fig.~\ref{C7F3} shows the case of a left-handed chiral host medium with $\kappa=0.003.$ 
As the handedness of the medium is opposite to the handedness of the incident trapping beam, the OT is always negative. Although the peak value is slightly smaller than 
the magnitude of the peak for $\kappa=-0.003,$
it still corresponds to a remarkable enhancement of  the 
negative OT effect when compared with the experiment reported in \cite{Diniz2019}. 
 
\begin{figure}\centering
\includegraphics[width = 3.4in]{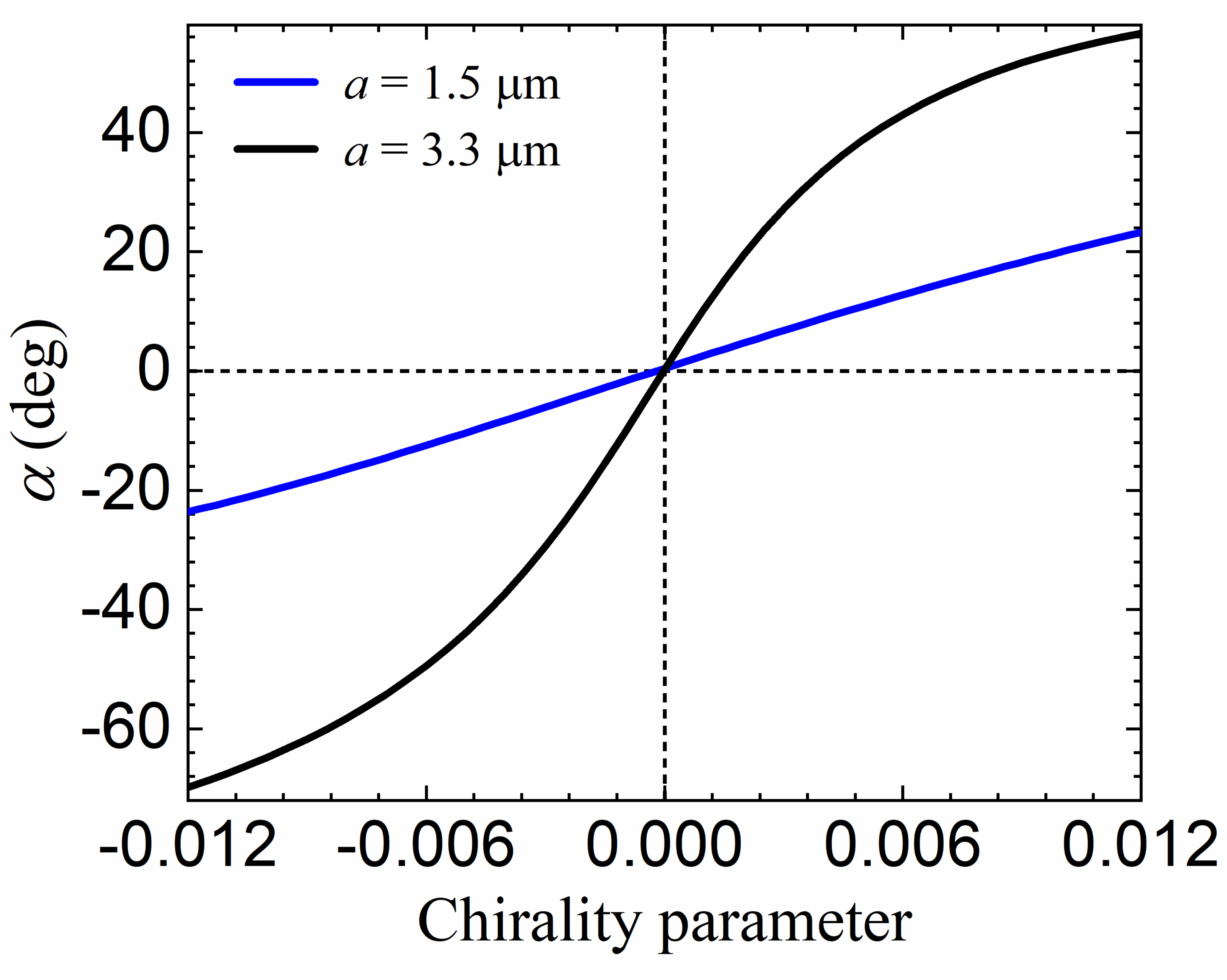}
\caption{Microsphere rotation angle $\alpha$ in degrees versus  chirality parameter $\kappa$ of the host medium for a silica microsphere of
 radius   $a=1.5 \,\mu{\rm m}$ (blue) and  $3.3\,\mu{\rm m}$ (black). The incident beam is right-handed circularly-polarized (helicity $\sigma=-1$).}
 \label{C7F4}
\end{figure}

The comparison between the results for opposite signs of $\kappa$ shown in 
Fig.~\ref{C7F3}
 shows that the sense of rotation can be employed as a direct 
probe of the handedness of the medium when using microspheres of radii $a>1\,\mu{\rm m}.$ 
The angle $\alpha$ indeed changes sign as $\kappa$ changes from negative to positive values as illustrated by Fig.~\ref{C7F4}, where 
we plot the rotation angle $\alpha$ as a function of the chirality parameter $\kappa$ for 
$a=1.5 \,\mu{\rm m}$ (blue) and  $3.3\,\mu{\rm m}$ (black).  

The strong dependence of the rotation angle on the chirality parameter illustrated by 
Fig.~\ref{C7F4} paves the way for 
an all-optical, local characterization of the host medium chiral response at the nanoscale with the help of optical tweezers. 
For the radius $a=3.3\,\mu{\rm m},$
 the slope of the function $\alpha(\kappa)$ in the neighborhood of $\kappa=0$ is $\Delta \alpha/\Delta\kappa \approx 9.9\times 10^3\,{\rm deg},$ thus allowing for a 
 chirality resolution $\delta \kappa \sim 2\times10^{-5}$ given a typical experimental precision $\delta \alpha \sim 0.2^{\rm o}.$
Such figures bring naturally occurring chiral solutions within reach of our proposal for characterization of chirality,
 which seems to be ideally suited for the small-volume microfluidic chambers 
 often employed as samples in optical tweezers setups~\cite{Marago2013}.

\section*{Discussion} \label{conclusions_}

 We have shown that the optical force acting on a dielectric trapped microsphere embedded in a chiral medium 
 strongly depends on the chirality parameter $\kappa$ and on the handedness of the CP trapping beam. 
 The trap axial stability is greatly enhanced by choosing a CP beam with the same handedness of the host medium. 
Such arrangement allows for optical tweezing of high-refractive index particles 
that cannot be trapped otherwise, { thus enlarging the scope of single-beam optical traps.}

We have also considered the optical torque on the trapped particle's center of mass, which 
 is characterized by the torsion constant $k_\phi,$
 in order to unveil the remarkable interplay between chirality and the transfer of optical spin angular momentum. 
 Our approach allows 
 for a clear identification of the spin-orbit contribution to the optical force and torque. Mie scattering 
 of a CP incident field gives rise to a field component with the reserved SAM~\cite{Schwartz2006,Haefner2009}. 
 Such spin-to-orbit conversion becomes more transparent in the formalism developed in this paper, 
as field components with opposite helicities propagate 
with different phase velocities in a chiral host medium. 
  The spin-orbit effect provides a 
 sizable contribution for radii of the order of the wavelength 
 and increases as the radius is decreased into the Rayleigh scattering regime. 

 The optical torque leads to a rotation of the equilibrium position when a lateral external Stokes force is applied.
 We have found a sizeable, detectable  enhancement of the rotation angle 
for radii $a\sim 3\,\mu{\rm m}$ for  media with chiral indices compatible with those of naturally occurring materials.
Since the angle depends strongly on the chirality parameter in this range of radii, 
one might characterize the local  chiral response
of small-volume samples typically employed in optical tweezers from measurements of the equilibrium position similar to those reported in Ref.~\cite{Diniz2019}.
In addition, the sense of rotation provides a direct indication of the handedness of the solution. 
Altogether our findings show that the torque in optical tweezers could be exploited as a novel all-optical method to locally probe the chiral response at the nanoscale. It is important to distinguish this method from traditional optical methods of enantioselection of chiral solutions, such as the rotatory power, which only apply for macroscopically large systems and can only provide an average chiral response.

When considering the torsion constant in a chiral host medium, the geometrical optics result is obtained only for 
radii much larger than usually required. 
We have obtained  interference oscillations which are typical
for radii larger than the wavelength~\cite{Mazolli2003}. They arise from an unusual 
 type of semiclassical Mie scattering near the focal region, with the leading contribution coming from small angular momenta (small multipole orders) \cite{Neto2000}.
We have found oscillation amplitudes 
much larger than the typical values for achiral materials \cite{Viana2007} when considering radii close to $a\sim 3\,\mu{\rm m}.$
Such oscillations open the way for the characterization of the microsphere diameter with nanometric precision. 
On the other hand, from a more fundamental perspective, it brings into light an unexpected feature of semiclassical Mie scattering \cite{Nussenzveig92}
 that requires further investigation. 

\section*{Methods}

\begin{figure}
\centering \bigskip
\includegraphics[width = 4.3in]{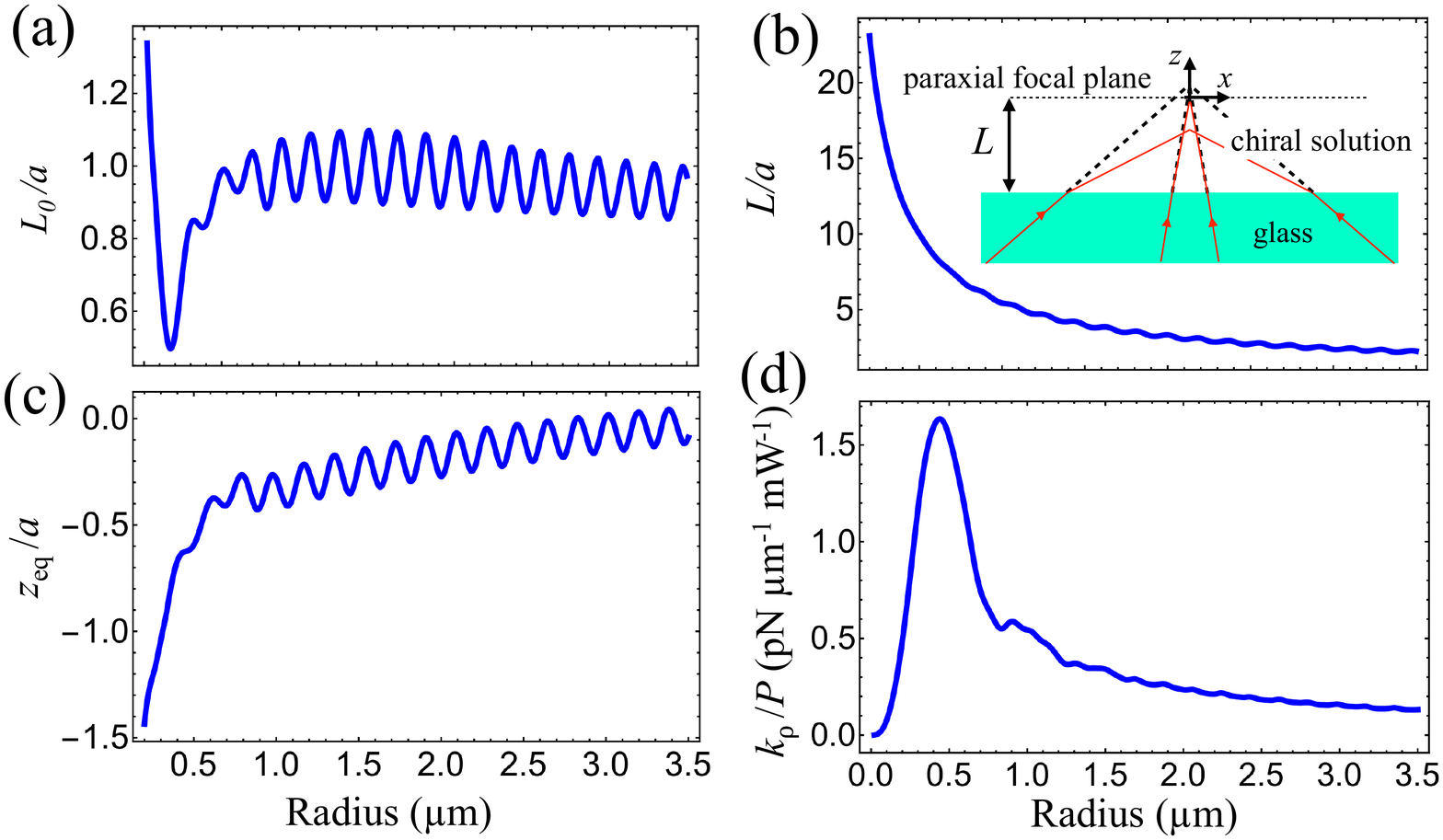}
 \caption{ Numerical simulation of the effect of refraction at the planar interface between the glass coverslip and the interior of the sample chamber. 
 As an example, we take $\kappa=-0.003.$
  (a) Initial reference position of the focal plane with respect to the glass slide versus radius. The reference configuration is defined by
  the condition that the equilibrium position is such that the microsphere is just touching the glass slide.  (b) Final focal plane position (in units of sphere radius)
  after displacing the objective by   $d= 5 \,\mu{\rm m}.$  (c)  Equilibrium position of the microsphere (in units of sphere radius) and (d)
   transverse trap  stiffness $k_{\rho}$ 
   (in units of power). }
\label{s1}
\end{figure}

\textbf{Numerical simulation of the spherical aberration introduced by the glass-sample interface.}
In real experiments, the distance $L$ between the paraxial focal plane and the planar interface between the glass coverslip and the sample chamber 
[see inset of Fig.~\ref{s1}(b)]
is not known beforehand. 
In order to control this parameter, which defines the amount of spherical aberration according to Eq.~(\ref{C7aberration_interface}), one can start from a reference 
configuration with the trapped microsphere just touching the coverslip, which is easy to identify experimentally~\cite{Viana2007}. 
Then, the objective is displaced away from the coverslip so as to 
trap the particle at a comfortable distance from the boundary of the chamber.

Here, we describe how we simulate such experimental procedure numerically. 
To illustrate the method, 
we show explicit intermediate results for 
silica microspheres embedded in a chiral solution with $\kappa = -0.003.$ All other examples discussed in the paper are obtained along the same lines. 

We first calculate the interface-focal plane distance $L_0$ in the  initial reference 
configuration, using the condition $z_{\rm eq}^{(0)}+L_0=a,$ where  $z_{\rm eq}^{(0)}$ is the particle equilibrium position (measured with respect to the focal plane) in the 
reference configuration. Thus, we solve $Q_z(z_p=a-L_0)=0$ for $L_0$ as a function of radius. The results are shown in Fig.~\ref{s1}(a). 

The second step is to increase $L$ by a controlled amount: $L=L_0+N_\sigma d,$ where $d$ is the objective displacement. In 
Fig.~\ref{s1}(b) we plot $L/a$ versus radius for $d=5\,\mu{\rm m}.$
 Once the coverslip-focal plane distance is known, 
 one can either directly compute the axial force as a function of the axial position $z_p$ (see Fig.~\ref{C7F2}), or 
 continue the procedure by solving $Q_z(z_p=z_{\rm eq})=0$ for the stable equilibrium position $z_{\rm eq}.$
  The results for $z_{\rm eq}$ as a function of radius,
 shown in Fig.~\ref{s1}(c),
 display interference oscillations with a characteristic period $\Delta a = \lambda_0/(4n_{\rm s})$ already discussed in connection with Fig.~\ref{kphi}. 

Finally, the last step consists in computing the partial-wave (multipole) series for $k_\phi$ and $k_\rho$ taking $\rho_p=0$ and $z_p=z_{\rm eq}.$ 
The results for $k_\phi/P$ and $k_\rho/P$ are shown in Fig.~\ref{kphi} (black line) and {Fig.}  \ref{s1}(d), respectively. Note that 
$k_\rho>0$ for all radii 
as required
for trap stability on the $xy$ plane.

\textbf{Multipole series for the radial and azimuthal force components.}
The axial components of the scattering and extinction optical force components, normalized by Eq.~(\ref{normalization}), are given by 
(\ref{C7Qszp}) and (\ref{C7Qez}), respectively. Here, we provide the remaining cylindrical components. 


\begin{itemize}
\item
 Scattering radial component
\begin{eqnarray}   \label{C7Qsrho}
&& Q_{s\rho}=-\frac{8\gamma^2}{F_{\sigma}N_{\sigma}} {\rm Im}\Biggl\{\sum_{\ell m}
 \frac{(2\ell+1)}{\ell(\ell+1)}\sqrt{(\ell-m)(\ell+m+1)}\,\sigma\,  \biggl[\biggl(1-\frac{\sigma\, \kappa}{n_{\rm m}}\biggr)\vert A_{\ell}\vert^2-\biggl(1+\frac{\sigma\, \kappa}{n_{\rm m}}\biggr)\vert B_{\ell}\vert^2\biggr]  \nonumber \\ &&  \times G^{(\sigma)}_{\ell,m}G^{(\sigma)*}_{\ell,m+1}
-
\frac{\sqrt{\ell(\ell+2)(\ell+m+1)(\ell+m+2)}}{\ell+1}
 \biggl[\biggl(1+\frac{\sigma\, \kappa}{n_{\rm m}}\biggr)A_{\ell}A_{\ell+1}^{*}+\biggl(1-\frac{\sigma\, \kappa}{n_{\rm m}}\biggr)B_{\ell}B_{\ell+1}^{*}\biggr]\nonumber\\ 
& & \times \left( G^{(\sigma)}_{\ell,m}G^{(\sigma)*}_{\ell+1,m+1} 
  +G^{(\sigma)}_{\ell,-m}G^{(\sigma)*}_{\ell+1,-(m+1)}\right)\Biggr\}
  \end{eqnarray}

\item
  Scattering azimuthal component
\begin{eqnarray} \label{C7Qsphi}
&&Q_{s\phi}= \frac{8\gamma^2}{F_{\sigma}N_{\sigma}}{\rm Re}\Biggl\{\sum_{\ell m }
 \frac{(2\ell+1)}{\ell(\ell+1)}\sqrt{(\ell-m)(\ell+m+1)}\,\sigma\,  \biggl[\biggl(1-\frac{\sigma\, \kappa}{n_{\rm m}}\biggr)\vert A_{\ell}\vert^2-\biggl(1+\frac{\sigma\, \kappa}{n_{\rm m}}\biggr)\vert B_{\ell}\vert^2\biggr] \nonumber \\ &&
 \times G^{(\sigma)}_{\ell,m}G^{(\sigma)*}_{\ell,m+1}
\nonumber
-
\frac{\sqrt{\ell(\ell+2)(\ell+m+1)(\ell+m+2)}}{\ell+1}
 \biggl[\biggl(1+\frac{\sigma\, \kappa}{n_{\rm m}}\biggr)A_{\ell}A_{\ell+1}^{*}+\biggl(1-\frac{\sigma\,\kappa}{n_{\rm m}}\biggr)B_{\ell}B_{\ell+1}^{*}\biggr]\\ & & \times
\left(G^{(\sigma)}_{\ell,m}G^{(\sigma)*}_{\ell+1,m+1} 
-G^{(\sigma)}_{\ell,-m}G^{(\sigma)*}_{\ell+1,-(m+1)}\right)\Biggr\}
\end{eqnarray}

\item
  Extinction radial component
\begin{equation}
Q_{e\rho}(\rho,\phi,z)=\frac{4\gamma^2} {F_{\sigma}N_{\sigma}}{\rm Im}\sum_{\ell m}(2\ell+1)
\biggl(1-\sigma\frac{(i-1)}{2}\frac{\kappa}{n_{\rm m}}\biggr) A_{\ell}\,  G^{(\sigma)}_{\ell,m} \,
\left(G^{(\sigma)-}_{\ell ,m+1} - G^{(\sigma)+}_{\ell,m-1}\right)^*
 \label{C7Qerho}
\end{equation}

\item
  Extinction azimuthal component
\begin{equation}
Q_{e\phi}(\rho,\phi,z)=-\frac{4\gamma^2} {F_{\sigma}N_{\sigma}}{\rm Re}\sum_{\ell m}(2\ell+1)
\biggl(1-\sigma\frac{(i-1)}{2}\frac{\kappa}{n_{\rm m}}\biggr) \,A_{\ell}\,G^{(\sigma)}_{\ell,m}\,
\left(G^{(\sigma)+}_{\ell,m-1}+G^{(\sigma)-}_{\ell,m+1}\right)^*
 \label{C7Qephi}
\end{equation}

\end{itemize}

The multipole coefficients 
  $G_{\ell m}^{(\sigma)}$  are given
 by Eq.~(\ref{C7Gjm}). 
The extinction radial and azimuthal components also require the coefficients
\begin{eqnarray}
G_{\ell,m}^{(\sigma)\pm}(\rho_p,z_p)=\int_{0}^{\theta_0}d\theta\sin\theta \sin\theta_{\rm m} \sqrt{\cos \theta}\,T(\theta)\, e^{-\gamma^2\sin^2\theta}d_{m\pm 1,\sigma}^{\ell}(\theta_{\rm m})\, J_{m-\sigma}\left( k_{\rm g} \rho_p\sin\theta \right)  e^{i[\Phi_{\rm sa}(\theta)+k_{\sigma}\cos\theta_{\rm m} z_p ]},
\end{eqnarray}

One can show that the above results, alongside Eqs.~(\ref{C7Qszp}) and (\ref{C7Qez}) for the axial components, are such that 
\[
(Q_z,Q_\rho,Q_\phi)\rightarrow (Q_z,Q_\rho,-Q_\phi)
\]
when taking 
$(\sigma\rightarrow -\sigma,\, \kappa\rightarrow -\kappa)$
as expected for the cylindrical components of a polar vector.

\section*{Acknowledgements}

We thank Kain\~a Diniz, Bruno Pontes, {Diney Ether} and Nathan Viana for inspiring discussions.   This work has been supported by 
 the Brazilian agencies National Council for Scientific and Technological Development (CNPq), 
  Coordination for the Improvement of Higher Education Personnel (CAPES),  the National Institute of Science and Technology Complex Fluids (INCT-FCx),
and the Research Foundations of the States of Minas Gerais (FAPEMIG), Rio de Janeiro (FAPERJ) and S\~ao Paulo (FAPESP).

\section*{Author Contributions}

R. A. and R. S. D. developed the theoretical formalism and wrote numerical codes.  F. A. P., F. S. S. R. and P. A. M. N. coordinated the work. All authors discussed the results and co-wrote the manuscript.

\section*{Additional Information}

\textbf{Competing Interests:} The authors declare no competing interests.

\end{document}